\documentclass[letter,longauth]{aa} 

\usepackage{longtable,lscape}
\usepackage{amsmath}
\usepackage{amsfonts}
\usepackage{natbib}
\bibpunct{(}{)}{;}{a}{}{,}
\usepackage{graphicx}
\usepackage{multirow}

\usepackage{graphicx}
\usepackage{txfonts}
\usepackage{natbib}
\usepackage[colorlinks=true,citecolor=blue]{hyperref}
\usepackage{color}
\usepackage{xspace}
\usepackage{dcolumn}
\usepackage{longtable}
\usepackage{lscape}
\usepackage{soul}        

\usepackage{rotating}
\usepackage{afterpage}

\usepackage{txfonts}
\begin{document}

   \title{Belt(s) of debris resolved around the Sco-Cen star HIP~67497\thanks{Based on observations made with ESO Telescopes at the Paranal Observatory under programme ID 097.C-0060(A)}$^{,}$ \thanks{This work is based on data products produced at the SPHERE Data Center hosted at OSUG/IPAG, Grenoble.}}


   \author{M. Bonnefoy 
           \inst{1}
          \and
          J. Milli\inst{2} 
           \and
          F. M\'{e}nard\inst{1}
          \and
          A. Vigan\inst{3}
          \and
          A.-M. Lagrange\inst{1}
          \and
          P. Delorme\inst{1}
          \and
          A. Boccaletti\inst{4}
          \and
          C. Lazzoni\inst{5}
          \and
          R. Galicher\inst{4}
          \and
          S. Desidera\inst{5}
          \and
          G. Chauvin\inst{1}
          \and
          J.C. Augereau\inst{1}
          \and
          D. Mouillet\inst{1}
          \and
          C. Pinte\inst{1}
          \and
          G. van der Plas\inst{6}
          \and
          R. Gratton\inst{5}
          \and
          H. Beust\inst{1}
          \and
          J.L. Beuzit\inst{1}
}

\institute{Univ. Grenoble Alpes, IPAG, F-38000 Grenoble, France.  CNRS, IPAG, F-38000 Grenoble, France \\
\email{mickael.bonnefoy@obs.ujf-grenoble.fr}
\and
European Southern Observatory (ESO), Alonso de Córdova 3107, Vitacura, Casilla 19001, Santiago, Chile
\and
Aix Marseille Université, CNRS, Laboratoire d’Astrophysique de Marseille, UMR 7326, 13388, Marseille, France
\and
LESIA, Observatoire de Paris, CNRS, Université Paris Diderot, Université Pierre et Marie Curie, 5 place Jules Janssen, 92190 Meudon, France
\and
INAF-Osservatorio Astronomico di Padova, Vicolo dell’Osservatorio 5, Padova, Italy, 35122-I
\and
DAS, Universidad de Chile, camino el observatorio 1515 Santiago, Chile}

   \date{Received May 13, 2016; accepted September 19, 2016}

 
  \abstract
   {}
  {In 2015, we initiated a survey of Scorpius-Centaurus A-F stars that are predicted to host warm-inner  and cold-outer belts of debris similar to the case of the system HR~8799. The survey aims to resolve the disks and detect planets responsible for the disk morphology.  In this paper, we study the F-type star HIP~67497 and present a first-order modelisation of the disk in order to derive its main properties.}
   {We used the near-infrared integral field spectrograph (IFS) and dual-band imager IRDIS of VLT/SPHERE to obtain angular-differential imaging observations of the circumstellar environnement of HIP~67497. We removed the stellar halo with PCA and TLOCI algorithms. We modeled the disk emission with the \texttt{GRaTeR} code.}
   {We resolve a ring-like structure that extends up to $\sim$450 mas ($\sim$50 au) from the star in the IRDIS and IFS data. It is best reproduced by models of a non-eccentric ring with an inclination of $80\pm1^{\circ}$, a position angle of $-93\pm1^{\circ}$, and a semi-major axis of $59\pm3$ au. We also detect an 	additional, but fainter, arc-like structure with a larger extension (0.65 arcsec) South of the ring that we model as a second belt of debris at $\sim$130 au.  We detect 10 candidate companions at  separations $\geq$1". We estimate the mass of  putative perturbers responsible for the disk morphology and compare it to our detection limits.  Additional data are needed to find those perturbers,  and to relate our images to large-scale structures seen with HST/STIS.}
  {}


   \maketitle
%

\section{Introduction}

The Scorpius-Centaurus OB association (Sco-Cen) is the nearest site of recent massive star formation \citep{1999AJ....117..354D}.  The proximity (d=90-200 pc) and young age ($\sim$11-17 Myr) of Sco-Cen all contribute to make it an excellent niche for direct imaging (DI) search of planets and disks. The planet imager instruments GPI \citep{2008SPIE.7015E..18M} and SPHERE \citep{2008SPIE.7014E..18B} have initiated surveys of the circumstellar environment of a few Sco-Cen stars at unprecedented contrasts and angular resolution. They have already resolved several new debris disks around those stars \citep[]{2015ApJ...807L...7C, 2015ApJ...812L..33K, 2015ApJ...814...32K, 2016A&A...586L...8L, 2016arXiv160502771D} with wing-tilt and ringed morphologies indicative of the presence of planets. 

We initiated a DI survey with SPHERE to image new giant planets and circumstellar disks around a sample of Sco-Cen F5-A0 stars with infrared excess. The excesses  can be modeled by  2 black-body components, each corresponding to a belt of debris \citep[][hereafter C14]{2014ApJS..211...25C}. This architecture is reminiscent of the Solar System, and of benchmark systems previously identified by DI like HR~8799 \citep{2008Sci...322.1348M, 2010Natur.468.1080M}, or  HD~95086 \citep{2013ApJ...779L..26R} which is also part of Sco-Cen.\\
In the course of the survey, we resolved a disk around the F0-type star HIP~67497 (HD~120326). This intermediate-mass star (M$=1.6 M_{\odot}$) is located at a distance of  $107.4^{+10.4}_{- 8.7}$ pc \citep{2007A&A...474..653V} and belongs to the 16 Myr  old \citep{2002AJ....124.1670M} Upper Centaurus Lupus sub-group \citep{1999AJ....117..354D}.  C14 modeled the infrared excess of the star with two belts of debris: a cold ($127\pm5$ K) bright ($L_{IR}/L_{*}=1.1\times10^{-3}$) belt at 13.9 au and a second colder ($63\pm5$ K) dimmer ($L_{IR}/L_{*}=1.4\times 10^{-4}$)  belt at 116.5 au. The same team produced an alternative model of the excess requiring only one belt located at 8.82$\pm$1 au \citep{2015ApJ...808..167J}. Previous observations with adaptive optics systems did not reveal any companion or structure close to HIP~67497 \citep{2013ApJ...773..170J}. Our resolved observations therefore offer the opportunity to better constrain the radial distribution of the dust around this star and to look for the companions responsible for that distribution. 

We present the observations and data in \S~\ref{section:data}, and the models in \S~\ref{section:diskprop}. We discuss the morphology of the disk and the existence of putative perturbers (planets) in \S~\ref{section:conclusion}.

\section{Observations and data reduction}
\label{section:data}

\begin{table*}
\caption{Log of observations,  April 6, 2016.}
\label{tab:obs}
\begin{center}
\begin{tabular}{lllllllll}
\hline\hline
			&					&	IRDIS &	 IFS & 	& &	&	&	\\
			\hline
UT time	&	Neutral density  &  \multicolumn{2}{c}{$\mathrm{DIT\times NDIT \times N_{EXP}}$}  &  $\Delta$PA & $<$Seeing$>$ &  Airmass	&	$\tau_{0}$ & Remarks	\\
HH:MM	&				&	(s)  & (s) & ($^{\circ}$)		&	(")		& & (ms) & 	\\
\hline
04:49	&	ND\_1.0    & 4$\times$2$\times$1  &   8$\times$1$\times$1 &  0.1 &  1.2 & 1.1 	& 3.7   & Unsat. PSF \\
04:50    &	ND\_0.0   & 32$\times$1$\times$1   &  64$\times$1$\times$1 & 	 0.1 	&	1.2 & 1.1	&	4.6	& Star center \\	
04:50    &	ND\_0.0   & 32$\times$8$\times$16  &  64$\times$4$\times$16  & 	 37.5  	&	1.1	& 1.1 & 3.5 & ADI sequence \\	
06:01	&	ND\_1.0    & 4$\times$2$\times$1 &   8$\times$1$\times$1 & 0.1 &  1.1   & 1.1 & 2.9 &  Unsat. PSF \\
06:02    &	ND\_0.0   & 32$\times$1$\times$1  &  64$\times$1$\times$1  &	0.1 	&	 1.0	&	1.1	&	3.0	& Star center \\	
\hline
\end{tabular}
\end{center}
\tablefoot{The seeing is measured at 0.5~$\muup$m. DIT {(Detector Integration Time)} refers to the individual exposure time per frame. NDIT is the number of individual frames per exposure, $N_{EXP}$ is the number of exposures, and $\Delta$PA to the amplitude of the parallactic rotation.}\\
\end{table*}

We observed HIP~67497  on April 6, 2016 with the VLT/SPHERE instrument \citep{2008SPIE.7014E..18B} operated in \texttt{IRDIFS} mode (Tab. \ref{tab:obs}). The mode enables for pupil-stabilized observations of the source placed behind an apodized Lyot coronagraph (92.5 mas radius) with the IRDIS \citep{2008SPIE.7014E..3LD} and IFS \citep{2008SPIE.7014E..3EC} sub-instruments operated in parallel.  IRDIS  yielded  images of $\sim$11.1$\times$12.4'' centered onto the star in the H2 ($\lambda_{c}$=1.593$\mu$m, width=52 nm) and H3 ($\lambda_{c}$=1.667$\mu$m, width=54 nm)  filters \citep{2010MNRAS.407...71V}. The IFS datacubes are made of 39 images covering the 0.96-1.33$\mu$m spectral range and a square field-of-view of 1.8 arcseconds. For registration purposes --- for both sub-instruments before and after the deep exposures --- we obtained frames with satellite spots (\textit{Star Center}) created by a waffle pattern introduced onto the deformable mirror of the instrument. Those frames were used to assess the position of the star behind the coronagraphic mask.  We recorded additional non-saturated exposures of the star placed outside of the coronagraphic mask with the neutral density filter \texttt{ND\_1.0} to estimate the flux and position of point sources.

\begin{figure}
\begin{center}
\includegraphics[width=\columnwidth]{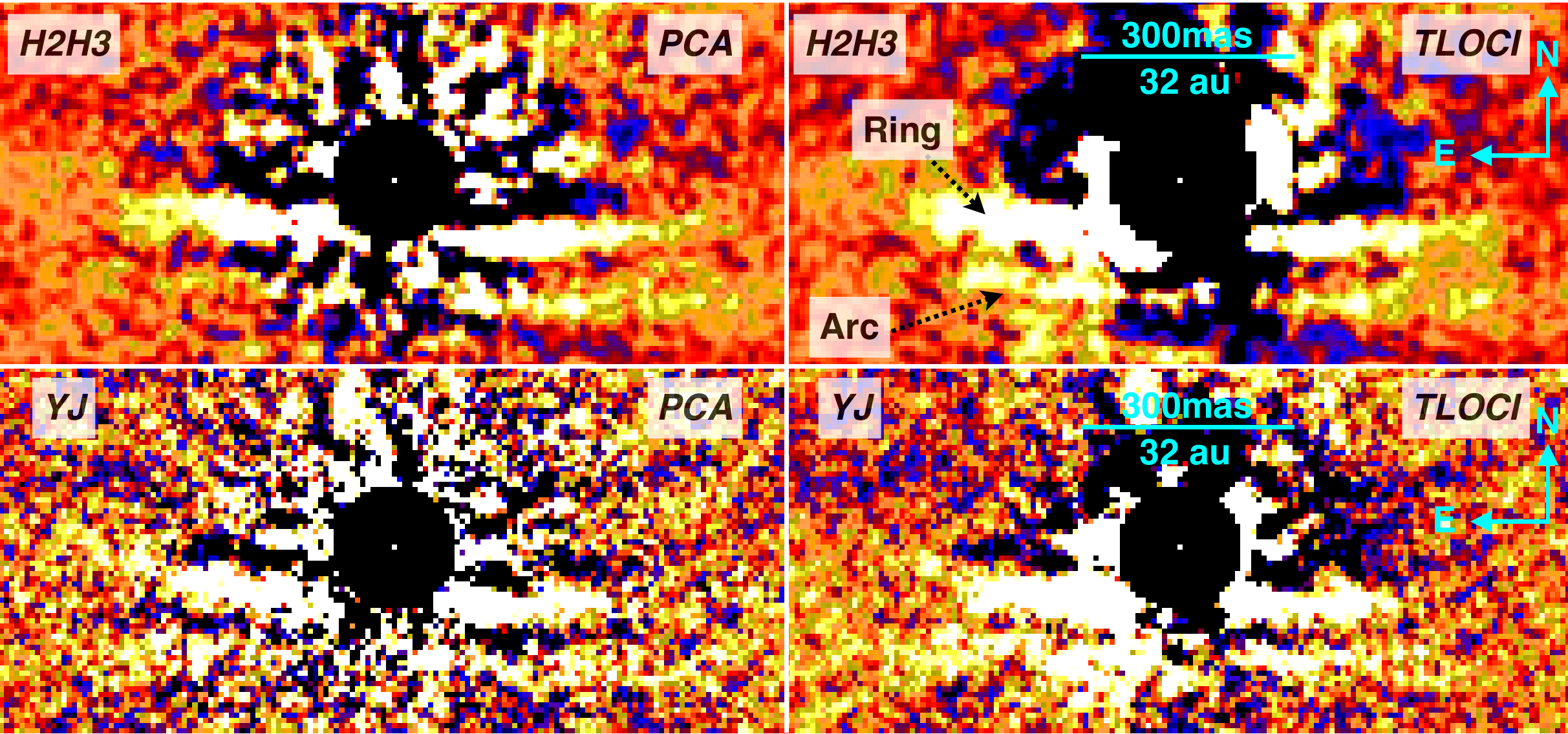}
\caption{Images of the disk structure around HIP~67497 obtained with the IFS (YJ), and IRDIS (H2H3), and considering PCA and TLOCI reductions.}
\label{fig:disk}
\end{center}
\end{figure}

We reduced the IRDIS data with the Data Reduction Handling software (DRH) of SPHERE \citep{2008ASPC..394..581P}.  The DRH corrected the raw science frames from the bad pixels, subtracted the dark, and recentered the images onto a common origin using the \textit{Star Center} images. The field rotation of  37.5$^{\circ}$ during the sequence of coronagraphic exposures was sufficient to apply the angular differential imaging (ADI) technique and reveal point sources down to 0.1".  We applied a  Principal Component Analysis (PCA) algorithm  \citep[][]{2012ApJ...755L..28S} on the H2 and H3 images separately to attenuate the stellar halo. The small number of modes kept in the PCA  (10) enabled to look for extended structures. We used  the TLOCI algorithm \citep{2014IAUS..299...48M} to identify point sources and to confirm the structures. 

The IFS data were analysed with a custom pipeline  \citep[][hereafter V15]{2015arXiv150900015V}. This tool subtracted the background,  interpolated the bad pixels, removed the instrument crosstalk, and performed a wavelength calibration of the spectra on the 2D raw frames. The DRH was then used to build the cubes from the 2D calibrated frames. We performed a PCA  taking advantage of the wavelength coverage and field rotation to look for point sources (V15). We also performed PCA and TLOCI reductions of the IFS data after collapsing each cube in wavelength in order to reveal the (extended, faint) disk structures. 

We used the True North value of $-1.649\pm0.019^{\circ}$, and platescales of $12.257\pm0.004$ and $7.46\pm0.02$ mas/pixel for IRDIS and the IFS, respectively. These values are derived from the observations of the NGC3603 astrometric field obtained as part of the SPHERE GTO survey on March 30, 2016 (priv. com.).

\begin{figure}
\begin{center}
\begin{tabular}{cc}
\includegraphics[width=4cm]{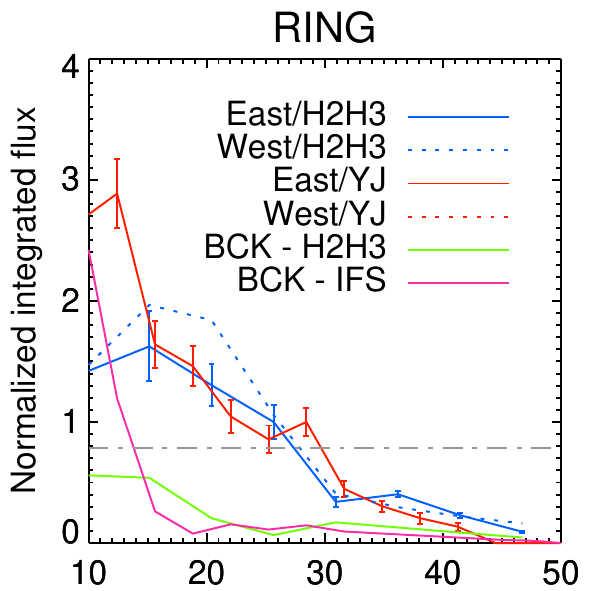} & \includegraphics[width=4cm]{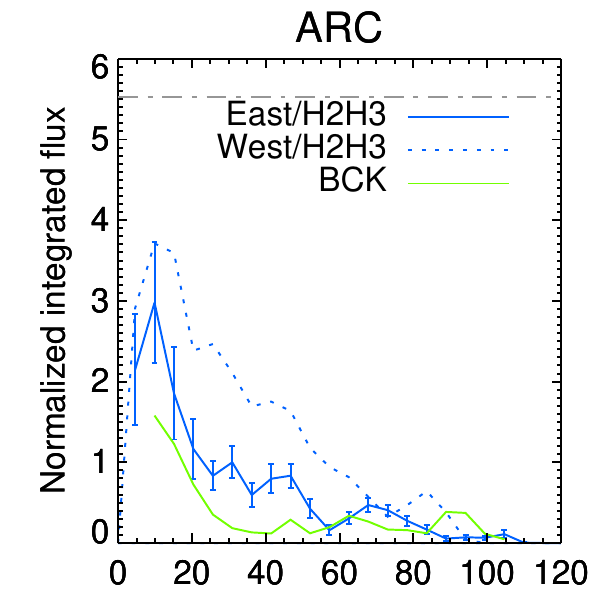}  \\
\includegraphics[width=4cm]{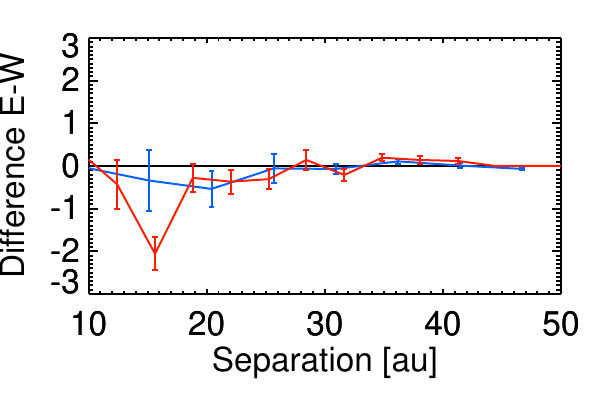} & \includegraphics[width=4cm]{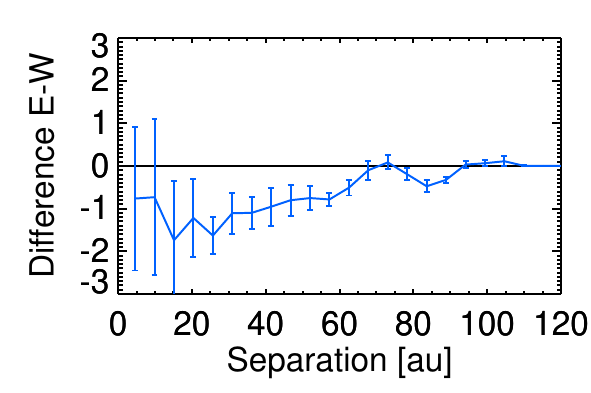}  \\
\end{tabular}
\caption{Radial profile along the disk position angle for the ring (left) and  for the arc (right), extracted from the H2H3, and YJ PCA images and normalized to the median emission of the east side from 15 to 40 au.  The residuals between the East and West disk profiles are shown at bottom. The normalization is different for the right and left panels. The dot-dashed line correspond to the cut of Fig. \ref{fig:disk}}
\label{fig:diskprofile}
\end{center}
\end{figure}

\section{Disk properties}
\label{section:diskprop}

\begin{figure*}
\begin{center}
\includegraphics[width=15cm]{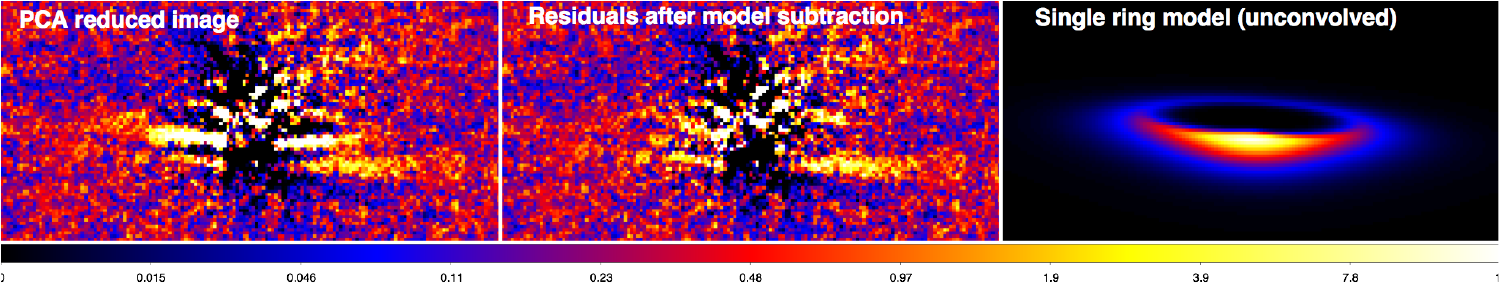}
\caption{Results of the forward modeling for the case of a single belt. Left: PCA image of the disk. Middle: Residuals after subtraction of the best single ring model, with the same linear color scale as the left image. Right: best single ring model (unconvolved with the SPHERE PSF) in a log scale to show the fainter backward side of the disk.}
\label{fig:diskmodel}
\end{center}
\end{figure*}

We detect an inclined ring at the same position in our collapsed YJ, H2, and H3 images (Fig. \ref{fig:disk}). 
The ring structure is seen both in our PCA and TLOCI reductions. It is also retrieved when exploiting both the ADI technique and the wavelength diversity of the IFS and IRDIS (ASDI).  We are  confident this is a real structure, not related to the stellar halo. The ring extends up to 450 mas, or 50 au, from the star in the IRDIS images. It has a smaller apparent extension (40 au) in the YJ images. This is likely due to the lower S/N of those data. We derived the flux profile \citep[non corrected from artifacts introduced by the ADI; ][]{2012A&A...545A.111M} along the position angle of the ring from the H2H3, and YJ PCA images (Fig. \ref{fig:diskprofile}). To improve the S/N, we averaged the flux over circular apertures of one FWHM diameter centered on that axis. The error bars for the East and West sides are computed from the dispersion of the flux within each aperture. They are of comparable amplitude.     We also report the flux profile of the residual background ("BCK") extracted  along the same appertures but tilted at a position angle  perpandicular to the disk one (green and pink lines). The profiles indicate that the ring is rougthly symmetric and has a similar flux profile in the YJ and H2/H3 bands from 15 to 50 au. Below 15 au and down to the coronagraph (10 au), the flux profile can be affected by residual speckles.

An additional arc-like structure ("Arc" in Fig. \ref{fig:disk}) is detected at 40 mas South of the main ring. It extends visually up to 650 mas on the west side (70 au). This arc structure shows a strong flux asymmetry compared to the ring (Fig. \ref{fig:diskprofile}). It is also retrieved at the same position into the IFS and IRDIS image for both TLOCI and PCA reductions. As for the ring, we are confident this feature is real and not a residual from the stellar halo. In the following section, we attempt a first characterization of the observed structures to better understand their nature. A more detailed modeling will be done in a forthcoming study.

\subsection{Model}
\label{sec_single_pop} 

We injected scattered light disk models into the IRDIS data to interpret the observed features. The disk models were generated with the \texttt{GRaTeR} code \citep{1999A&A...348..557A}. \texttt{GRaTeR} considers a power-law for the radial distribution of the dust of index $\alpha_\text{in}>0$ and $\alpha_\text{out}<0$ inward and outward of a reference radius $r_0(\theta)$ respectively.  The disk model is rotated to the angles of the initial frames, convolved with the measured PSF and then subtracted from the data. The resulting images are re-reduced using the  PCA algorithm used for the data reduction to obtain a disc-subtracted image. This step is repeated iteratively by varying the free parameters of the disk model until a reduced $\chi^2$ computed in the part of the model-subtracted image in a zone covering the location of the disk on the averaged H2+H3 images is minimized.  

 We assumed a grey color for the disk between H2 and H3. To limit the parameter space, we adopted a distribution of optically thin dust with constant effective scattering cross-section. 

\subsection{Inner ring}
\label{sec_inner_ring}

To model the inner ring, we fixed the inner and outer slope to $\alpha_\text{in}=10$ and $\alpha_\text{out}=-5$ to mimic sharp edges. We assumed a circular ring, with anisotropic scattering parametrized by a Henyey-Greenstein parameter g. Therefore, the model has five free parameters (semi-major axis $r_{0}$, position angle PA, inclination $i_{tilt}$, g,  and the flux). The $\chi^{2}$ (hereafter  $\chi^2_\text{region\:1}$) is computed in a region encompassing only the inner belt (see App. \ref{AppA}). The parameters of the best-fitting model are reported in Tab. \ref{tab_disc_param} and details on the minimization are given in App. \ref{AppA}. The initial PCA-reduced image, best model and PCA-reduced image after best model subtraction are shown in Fig. \ref{fig:diskmodel}.

 The "arc" structure is still present in the residual image (Fig. \ref{fig:diskmodel}), thus indicating that it is not a product of self-subtraction effects from the ring. We compare two possible models of that structure in the following section.

\begin{table}
\caption{Morphology of the inner belt}             
\label{tab_disc_param}      
\centering                          
\begin{tabular}{ccccc}        
\hline\hline                 
PA ($^{\circ}$) & $r_{0}$ (au) & $i_{tilt}$ ($^{\circ}$) & g & $\chi^2_\text{region\:1}$ \\    
\hline                        
-93$\pm1$ & $58.6\pm3$ & $80\pm1$ & 0.82$\pm0.02$ & 2.6 \\     
 \hline
 \end{tabular}
\end{table}

\subsection{Model of the arc}
\label{sec_outer_ring}

We considered the best model of the inner ring (\S~\ref{sec_inner_ring}) and modeled the arc with a second, outer belt. 
We assumed the same inclination, position angle and g parameter as for the inner belt.
The outer slope $\alpha_\text{out}$, the radius and the flux ratio with respect to the inner belt are left as free parameters. The $\chi^2$ (hereafter  $\chi^2_\text{region\:2}$) is now computed on a region encompassing the pixels where the ring and the arc are detected (App. \ref{AppA}). 

As an alternative, we investigated whether a narrow belt and a smooth disk halo could reproduce the ring and the arc. We did so by analogy with HR~8799  where such a halo was proposed \citep{2009ApJ...705..314S}.   We used for the ring the best fit found in  \S~ \ref{sec_inner_ring}, and added a smooth halo modeled by an additional disk of outer slope $\alpha_\text{out,halo}<\alpha_\text{out}$ and with a flux scaling factor lower than that of the narrow belt. In total, there are  3 free parameters: $\alpha_\text{out,halo}$, the narrow belt flux, and the halo flux.

\begin{table}
\caption{Results of the modeling of the belt and arc}             
\label{tab_disc_param2}      
\centering                          
\begin{tabular}{c c c c c}        
\hline \hline                 
Model	&	$\alpha_\text{out}$ & $r_{0}$ (au) & flux ratio & $\chi^2_\text{region\:2}$ \\    
\hline                        
Two belts	&  -8$^{+4}_{-8}$ & 130$\pm8$ & 0.08$\pm0.03$ & 3.5 \\     
Belt+halo & -1.3$^{+0.6}_{-0.8}$ & n.a.	&	0.66$\pm$0.40	&	3.8 \\
\hline                 
\end{tabular}
\end{table}

\begin{figure}
\begin{center}
\includegraphics[width=\columnwidth]{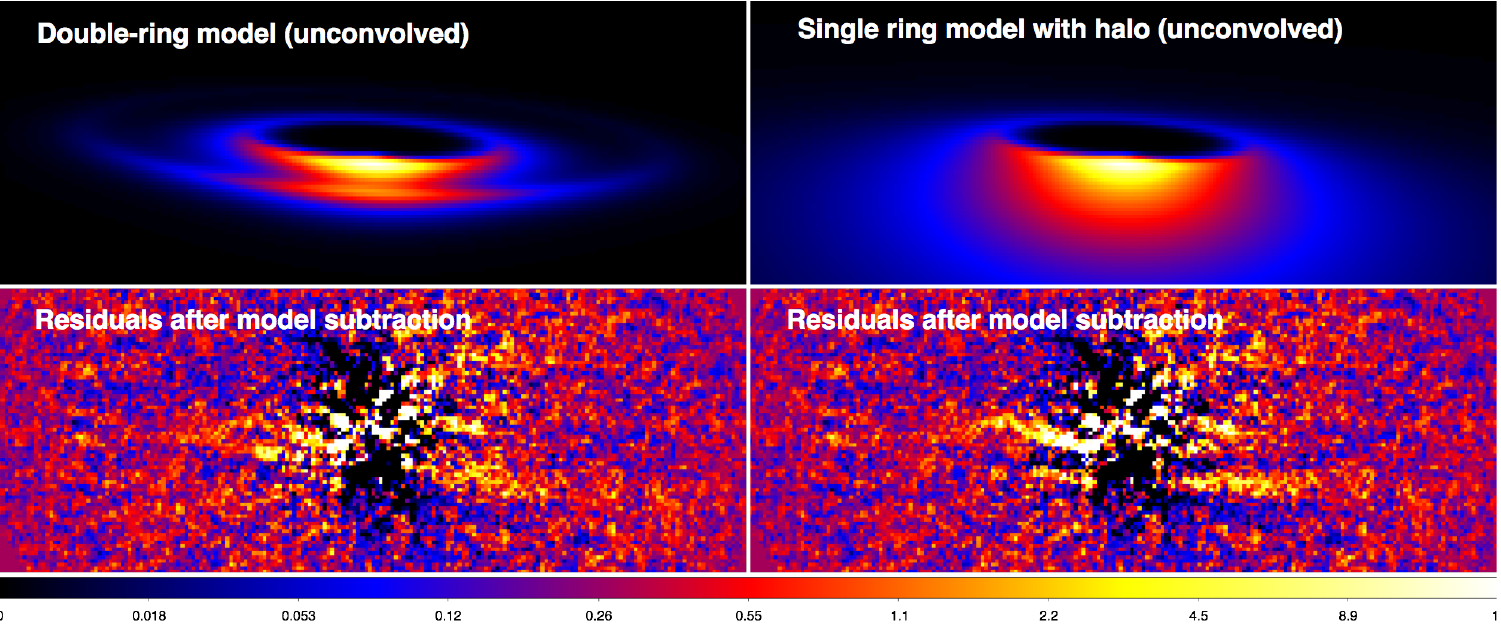}
\caption{Results of the forward modeling for the case of two belts (left panels, reduced $\chi^{2}=3.5$) and a narrow-ring laid over a smooth halo (right, reduced $\chi^{2}=3.8$). The models are shown on top and the post-ADI residual images at the bottom.}
\label{fig:diskmodel2}
\end{center}
\end{figure}

The best fit parameters are given in Tab. \ref{tab_disc_param2} and the residuals after subtraction of the two models are shown in Fig. \ref{fig:diskmodel2}. The residuals are slightly higher for the case of the ring+halo model. The scenario including two belts of debris is favored by the present data.

\section{Concluding remarks}
\label{section:conclusion}
The disk around HIP~67497 has an infrared luminosity (IL) in the same range as the disks already resolved around the Sco-Cen A-F type stars HD 111520, HIP 79977, HD 106906, and HD 115600 ($L_{IR}/L_{*}\sim10^{-3}-10^{-4}$). This could explain why we were able to resolve it. For HD~106906, and HD~115600, C14 finds outer belts at shorter separations than observed. In the case of HIP~67497, the second belt corresponds roughly to the location of the coldest belt found by  C14, but the flux ratio between the two belt models found in \S~\ref{sec_outer_ring}  is of the same order as the ratio of the IL of the belts found by C14 (7.85).

\begin{figure}
\begin{center}
\includegraphics[width=\columnwidth]{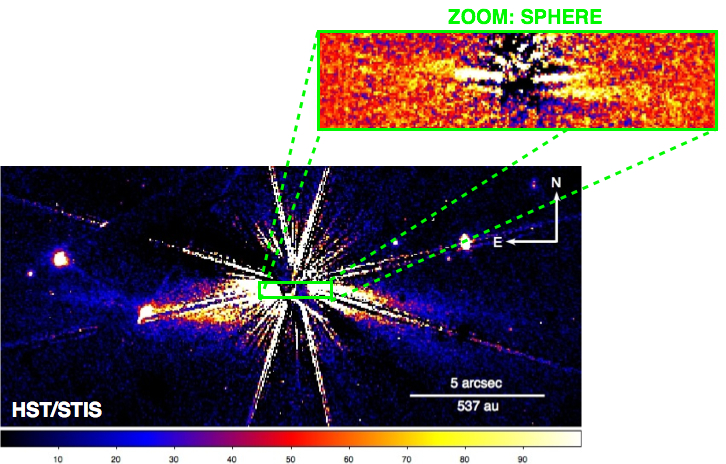}
\caption{Sketch showing the HST/STIS  and the SPHERE images of the debris disk around HIP~67497.}
\label{fig:HST}
\end{center}
\end{figure}

HST/STIS  images of HIP~67497 show extended emission at large scales \citep{2016IAUS..314..175P}, which we compare to the SPHERE images in Fig. \ref{fig:HST}. The  inner structures  ($\leq$3'') seen with HST have an orientation compatible with the belt and the arc of the SPHERE images. Another asymmetric feature is seen on the east side up to 8 arcsec, but can not be easily related to the rest. Different asymmetries at different scales have already been noticed for HD106906  \citep{2015ApJ...814...32K} or HD 32297 \citep{2014ApJ...780...25E, 2014AJ....148...59S} for instance. One candidate companions (\#1, see App. \ref{section:companions}) could lie within the structures revealed by STIS and the arc seen with SPHERE. The present STIS data are unfortunately affected by blind zones caused by the position of the coronagraphic bars of STIS (only two roll angles).

Several options exist to explain the morphology of the disk of HIP~67497.  The observed ring-like structures may be caused
 by dust-gas interactions \citep{2013Natur.499..184L}. Unfortunately, the gas content of the disk remains unknown.  Alternately, planets with different individual eccentricities
and semi-major axes \citep{2016ApJ...827..125L} may also provide an explanation for the double-ring structure. Our observations are sensitive to 1.5 to 15 M$_{Jup}$ in-between the  ring and the arc when accounting for the disk inclination (Fig. \ref{fig:dl}). We explore in App. \ref{AppC:perturbers} the case of one, two, or three perturbers on circular orbit, or one and two planets on eccentric orbits using numerical simulations. The predicted masses can reach $\sim$21 M$_{Jup}$ for the case of a single planet.  But the  mass estimate of the perturber(s) is sensitive to the eccentricity of the orbit(s). New observations with STIS and SPHERE are required to reveal the full morphology of the disk, improve the detection performances, and to clarify the nature of the CCs.

\begin{acknowledgements}
We thank our  referee for his/her constructive review of the manuscript. We also thank the ESO staff for gathering those data. We are grateful to the SPHERE consortium for providing  instrument platescale and True North. We  ackn. support in France from the ANR (grant ANR-14-CE33-0018), the PNP, and the PNPS. Gvdp ackn. support from the Millennium Science Initiative (grant RC130007), and from FONDECYT (grant 3140393).

\end{acknowledgements}

\bibliographystyle{aa}
\bibliography{HIP67497}

 \begin{appendix}
\section{Details on the model fitting}
\label{AppA}
\begin{figure}
\begin{center}
\includegraphics[width=\columnwidth]{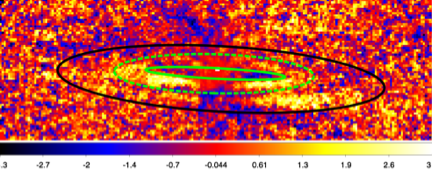}\\
\caption{Conservative SNR map showing the regions used to compute the chi square}.
\label{fig:chi2z}
\end{center}
\end{figure}

We report in Fig. \ref{fig:chi2z} the zones used to compute the $\chi^{2}$ when comparing the disk model to the data. The  $\chi^{2}$ for the single belt modelling is computed in the region between the plain green ellipse and the dashed green ellipse. It contains 1041 pixels, or 118 resolution elements. For the case of the model of the outer ring, the  $\chi^{2}$ is computed in the region between the plain green ellipse and the black ellipse. It contains 3279 pixels, or 390 resolution elements.

\begin{figure}
\begin{center}
\begin{tabular}{cc}
\includegraphics[width=4cm]{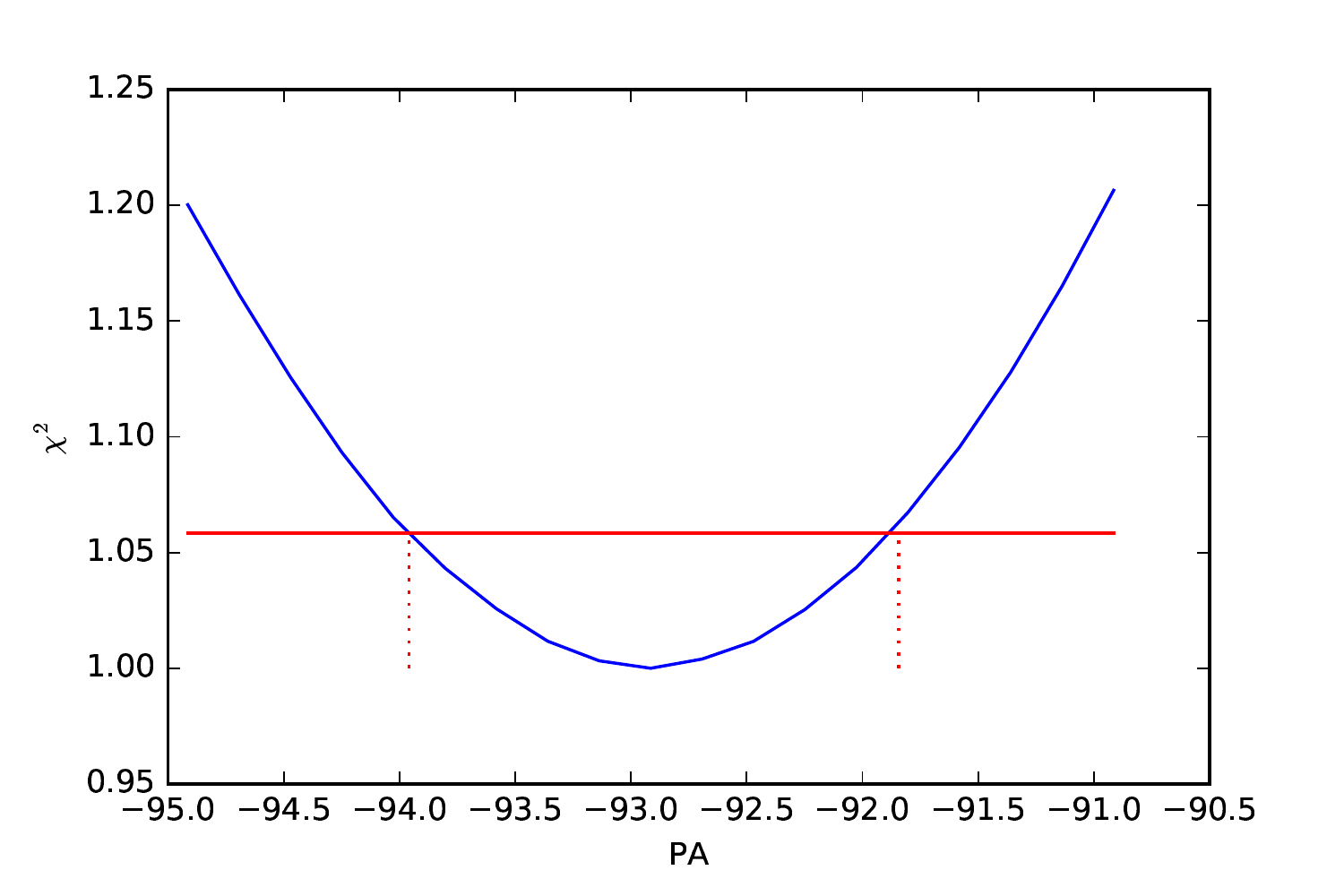} & \includegraphics[width=4cm]{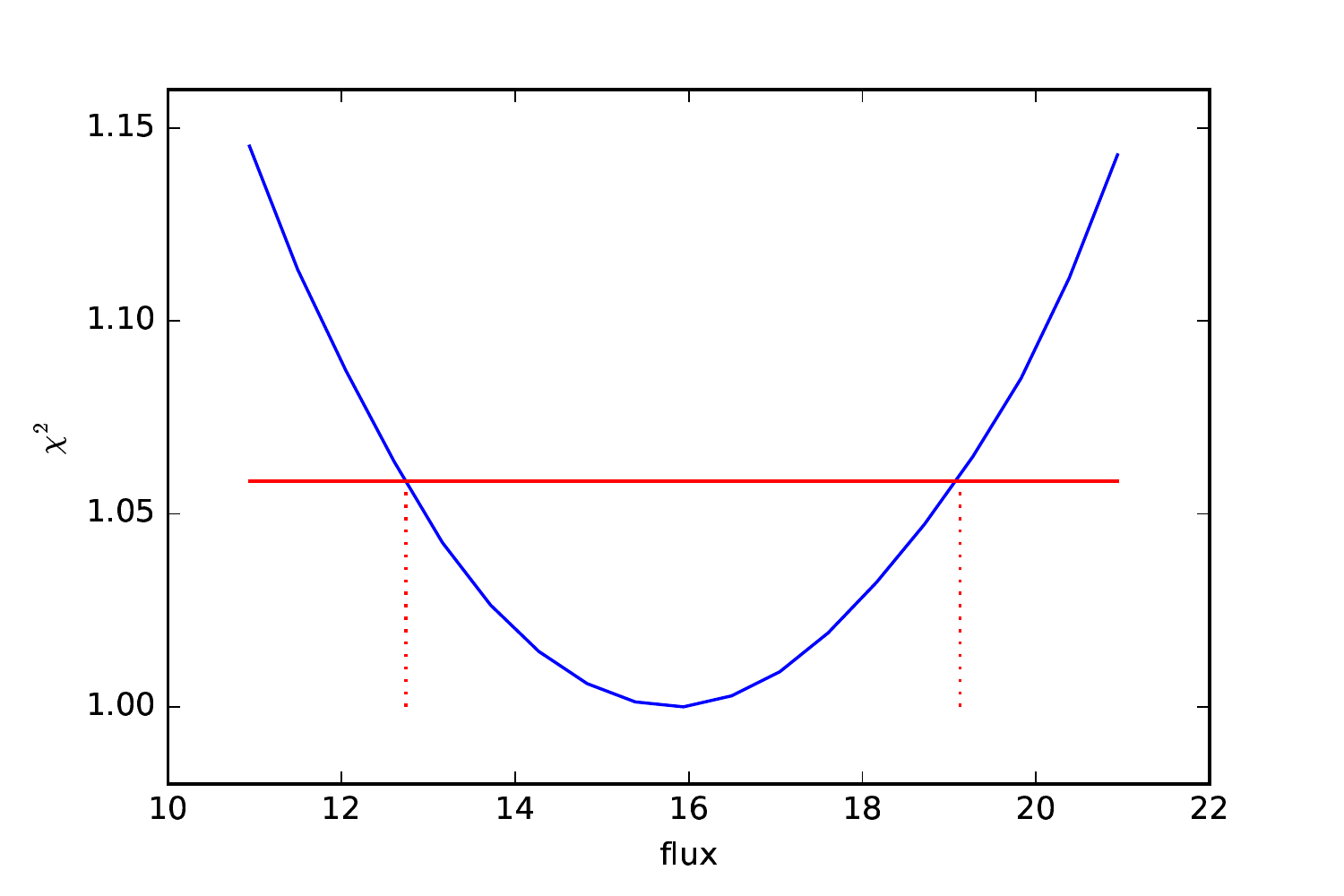} \\	
\includegraphics[width=4cm]{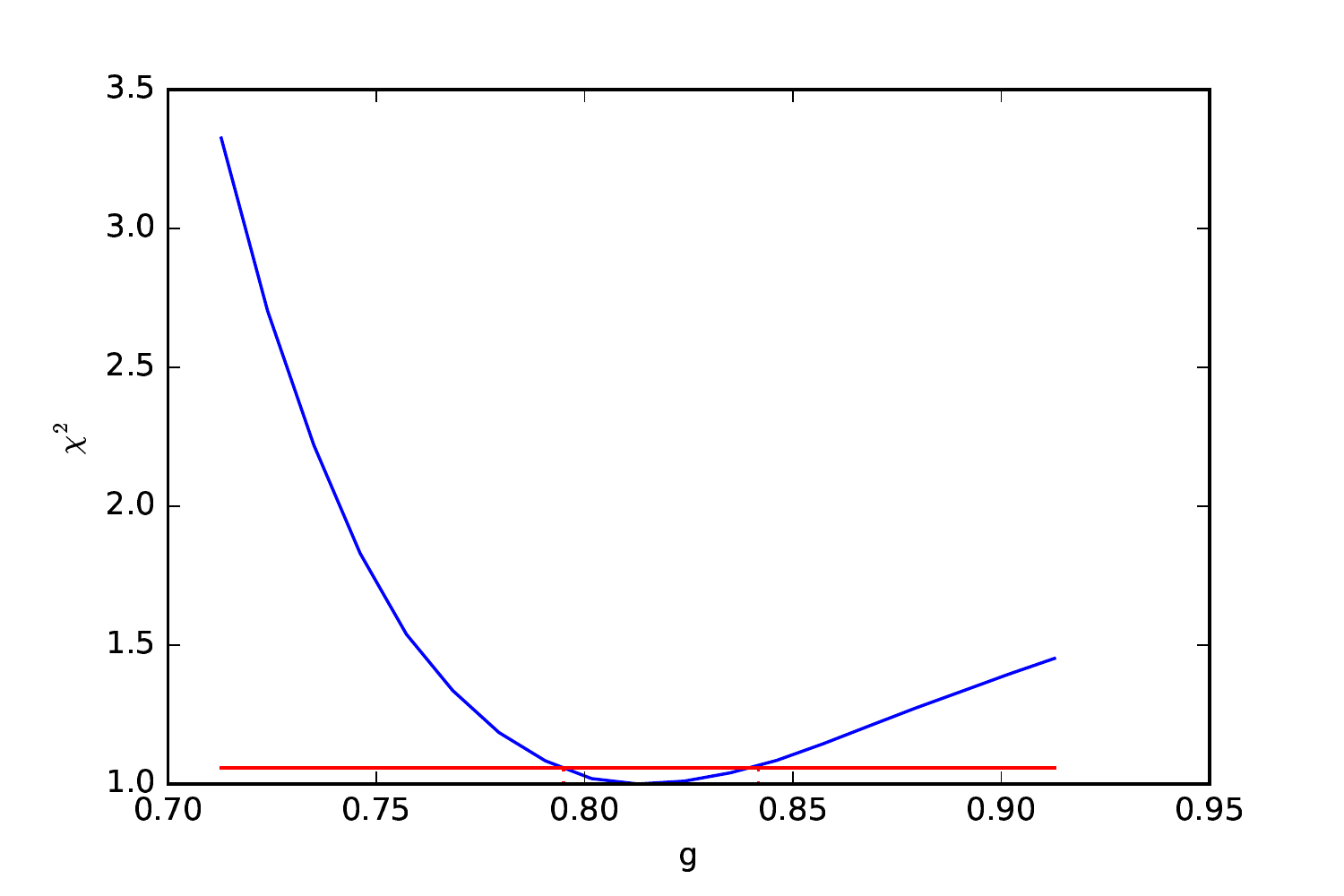} & \includegraphics[width=4cm]{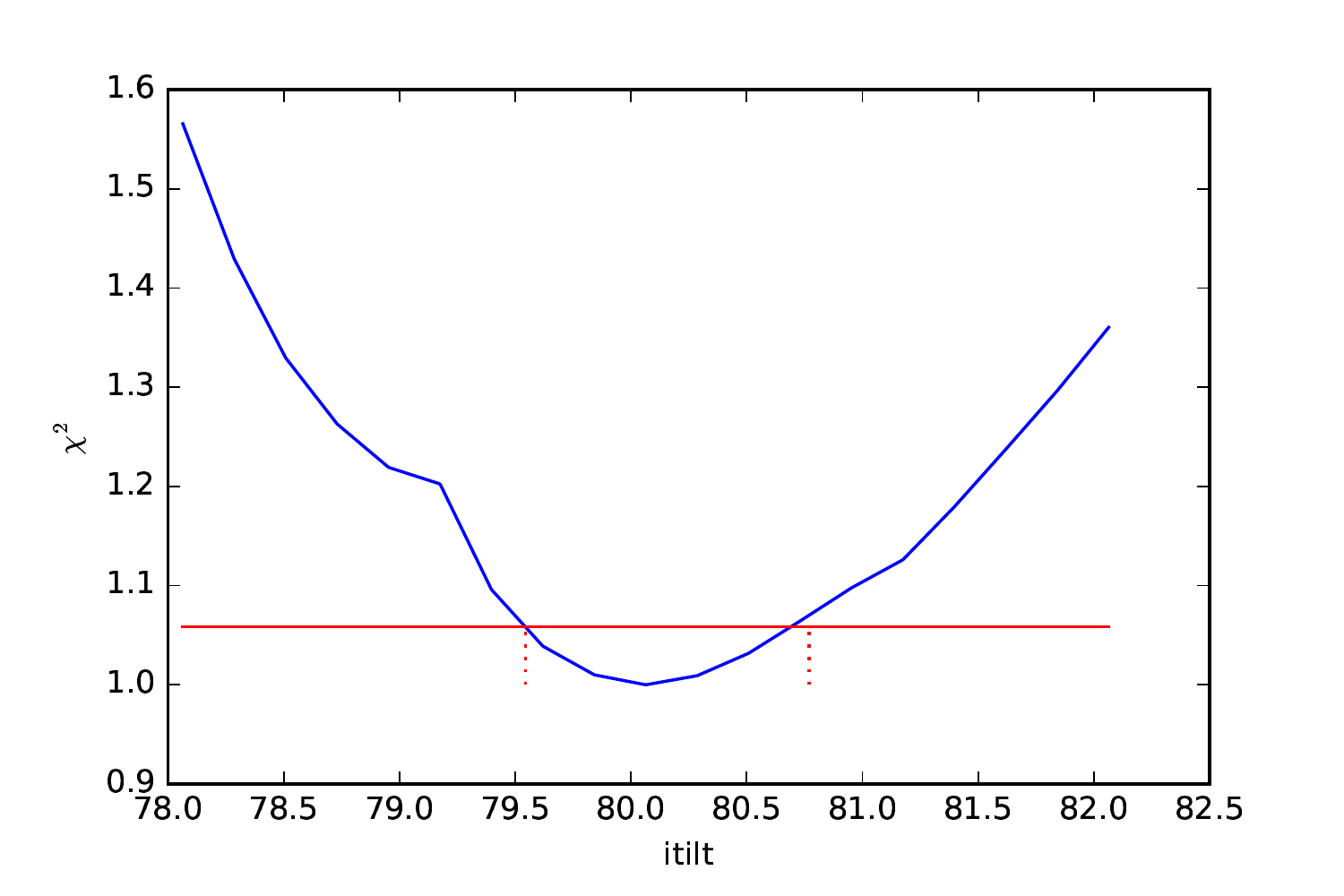} \\	
\includegraphics[width=4cm]{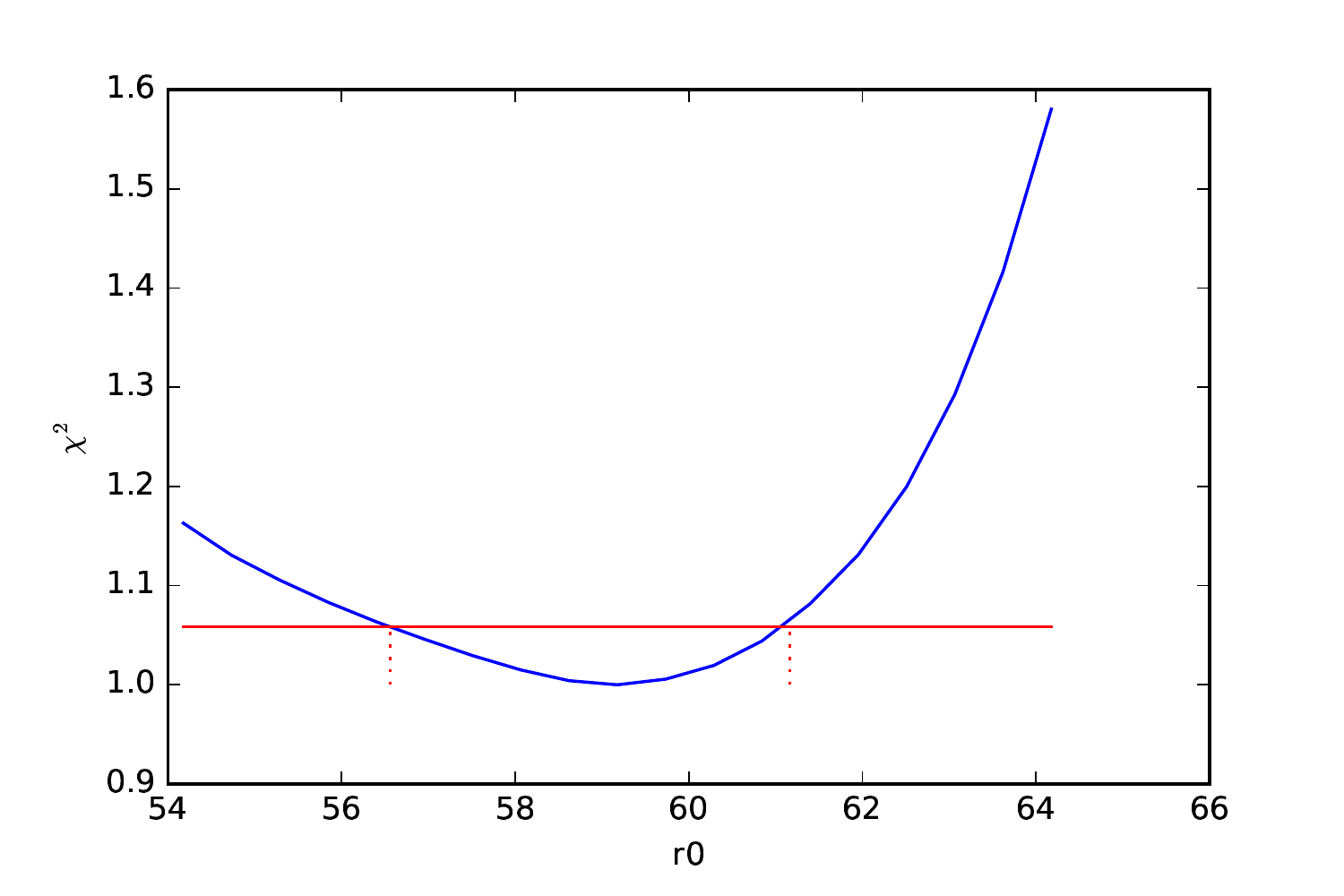} & \\
\end{tabular}
\caption{Minimization of the free parameters of the single belt model.}
\label{fig:modsing}
\end{center}
\end{figure}

\begin{figure}
\begin{center}
\begin{tabular}{cc}
\includegraphics[width=4cm]{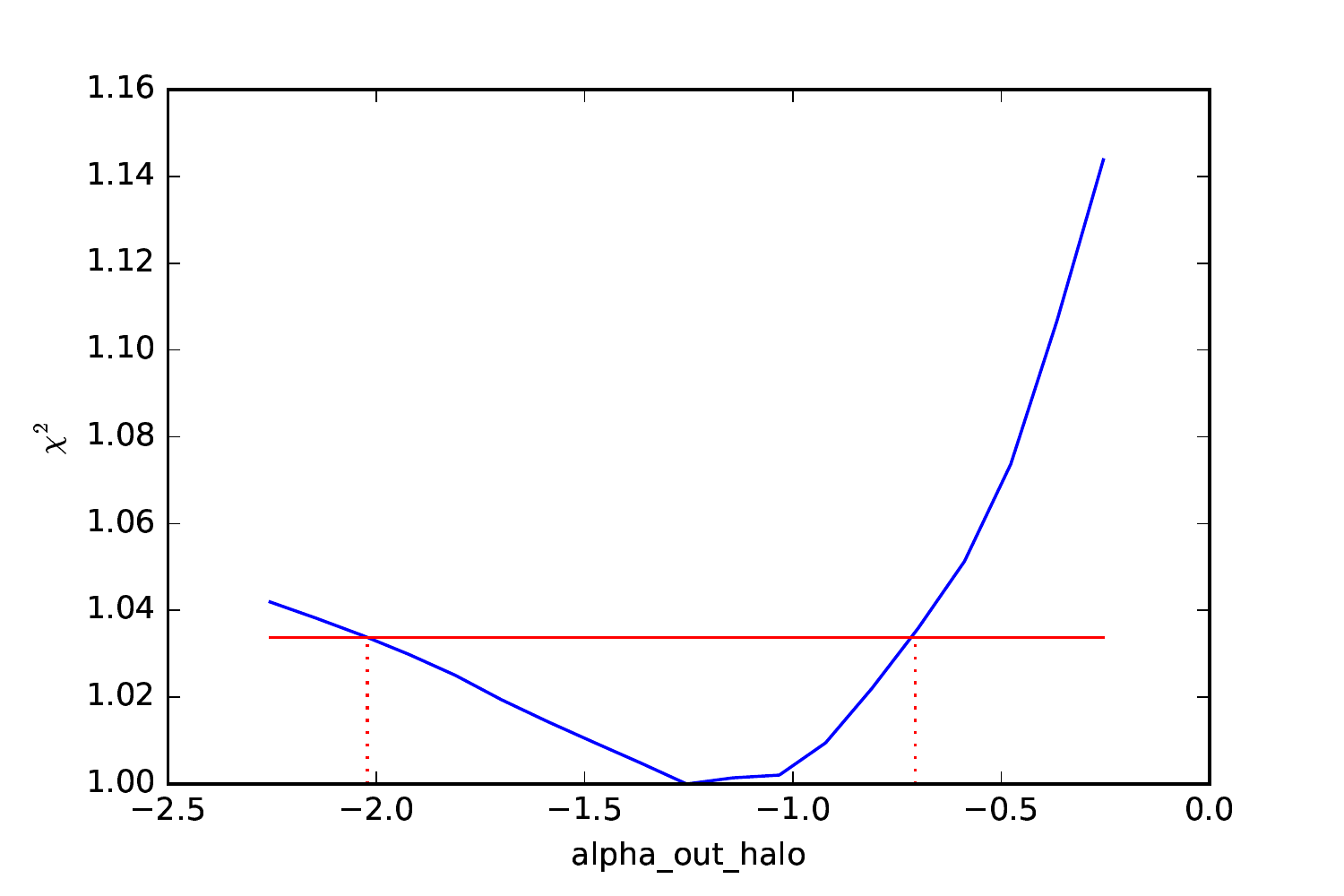} & \includegraphics[width=4cm]{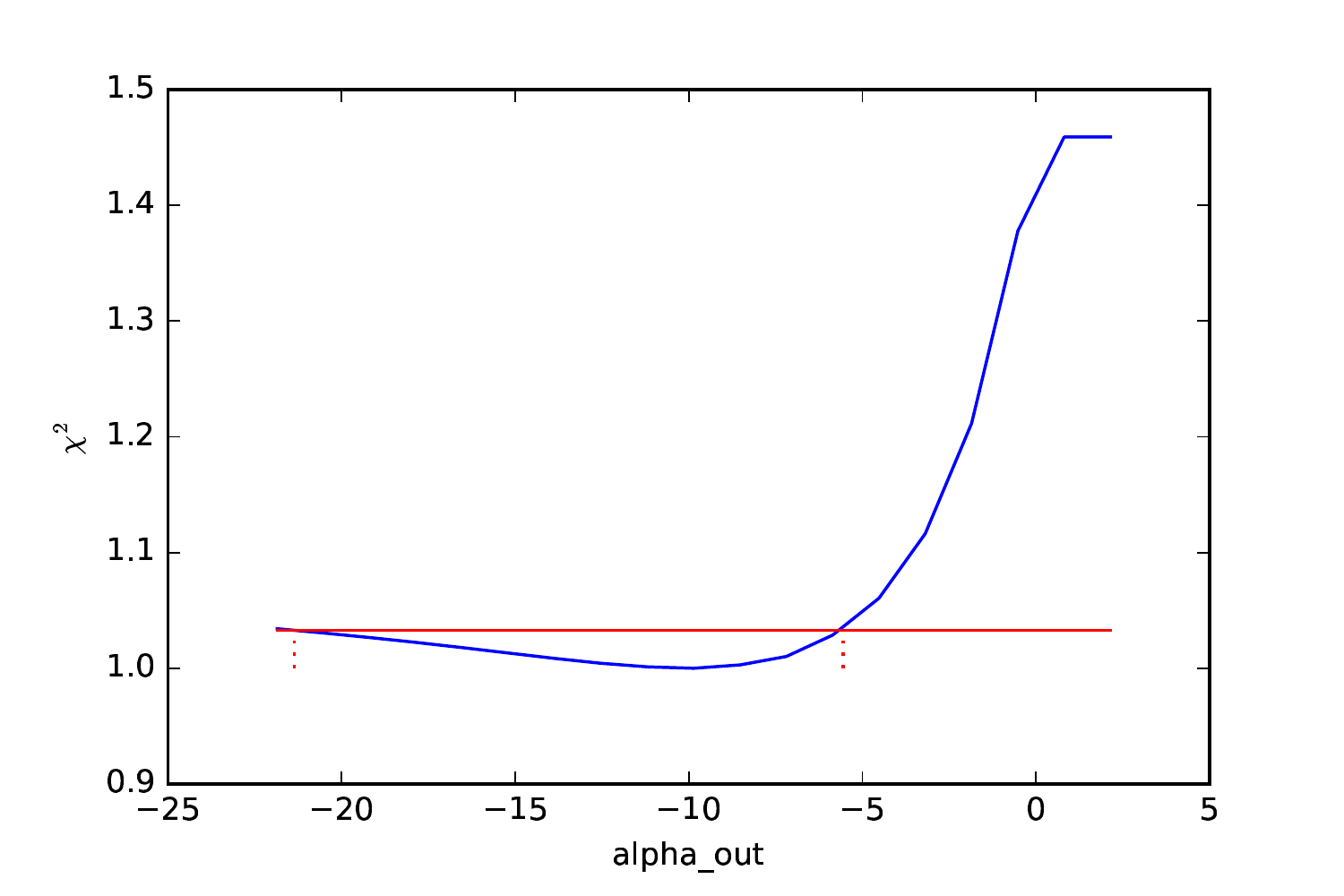} \\	
\includegraphics[width=4cm]{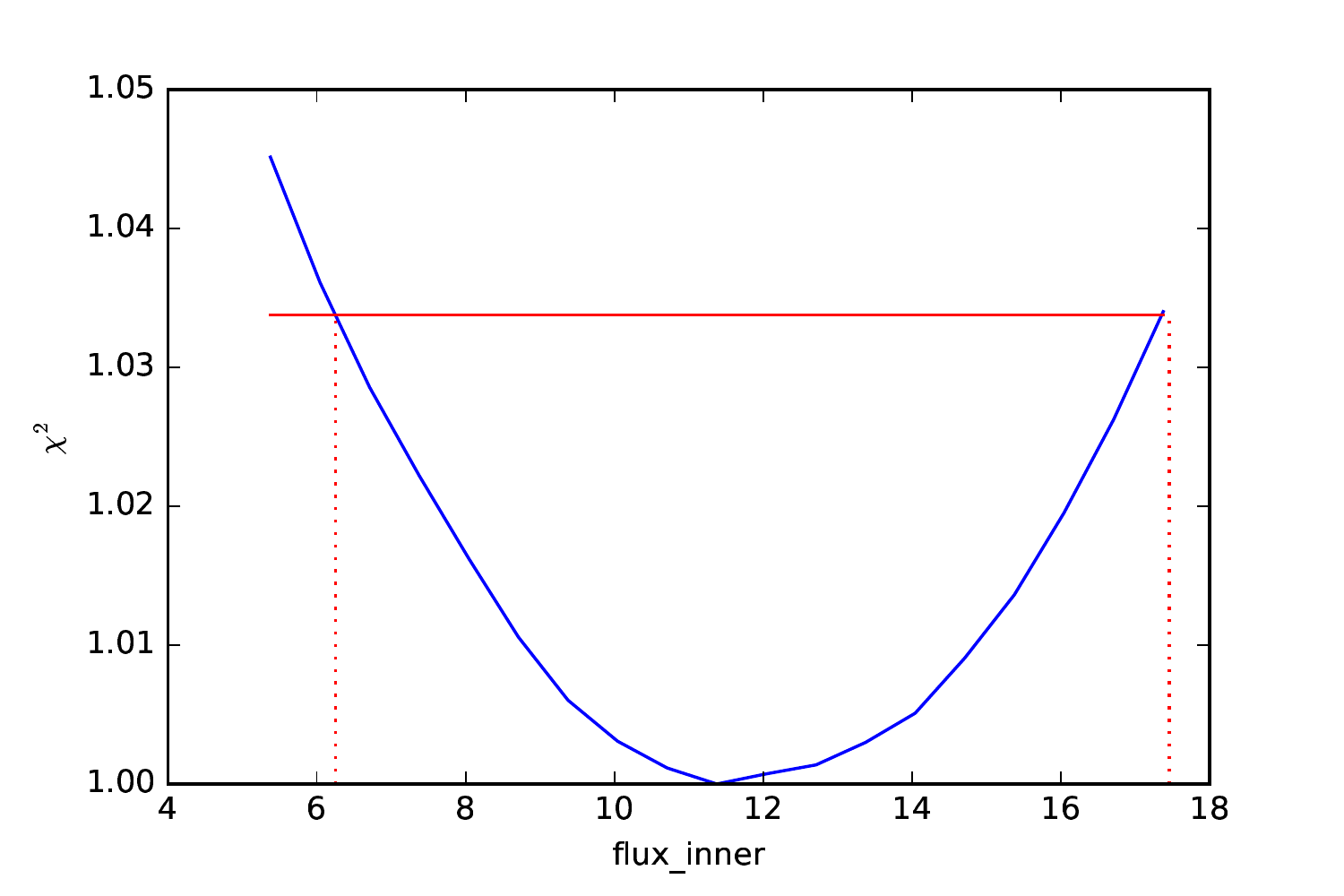} & \includegraphics[width=4cm]{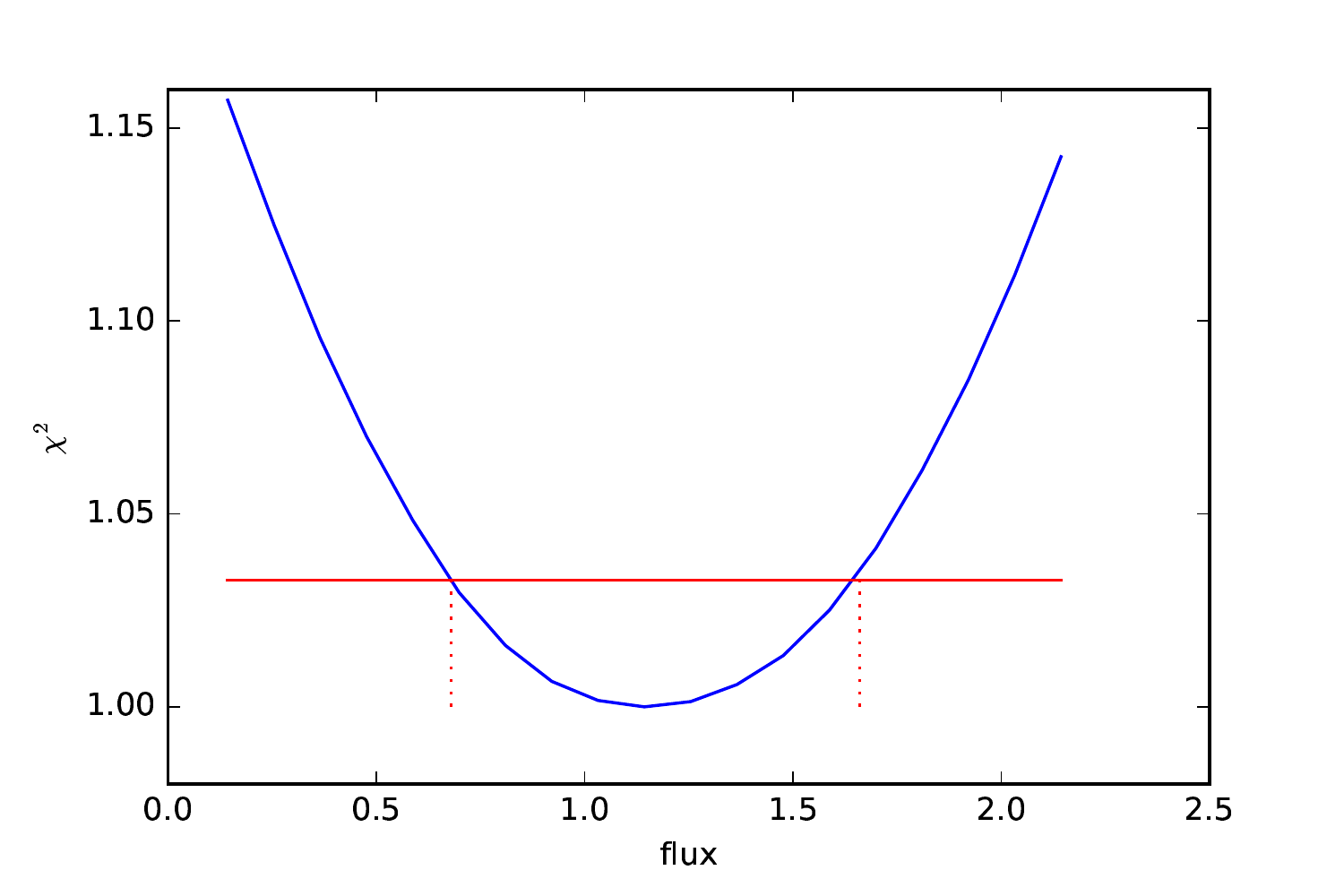} \\	
\includegraphics[width=4cm]{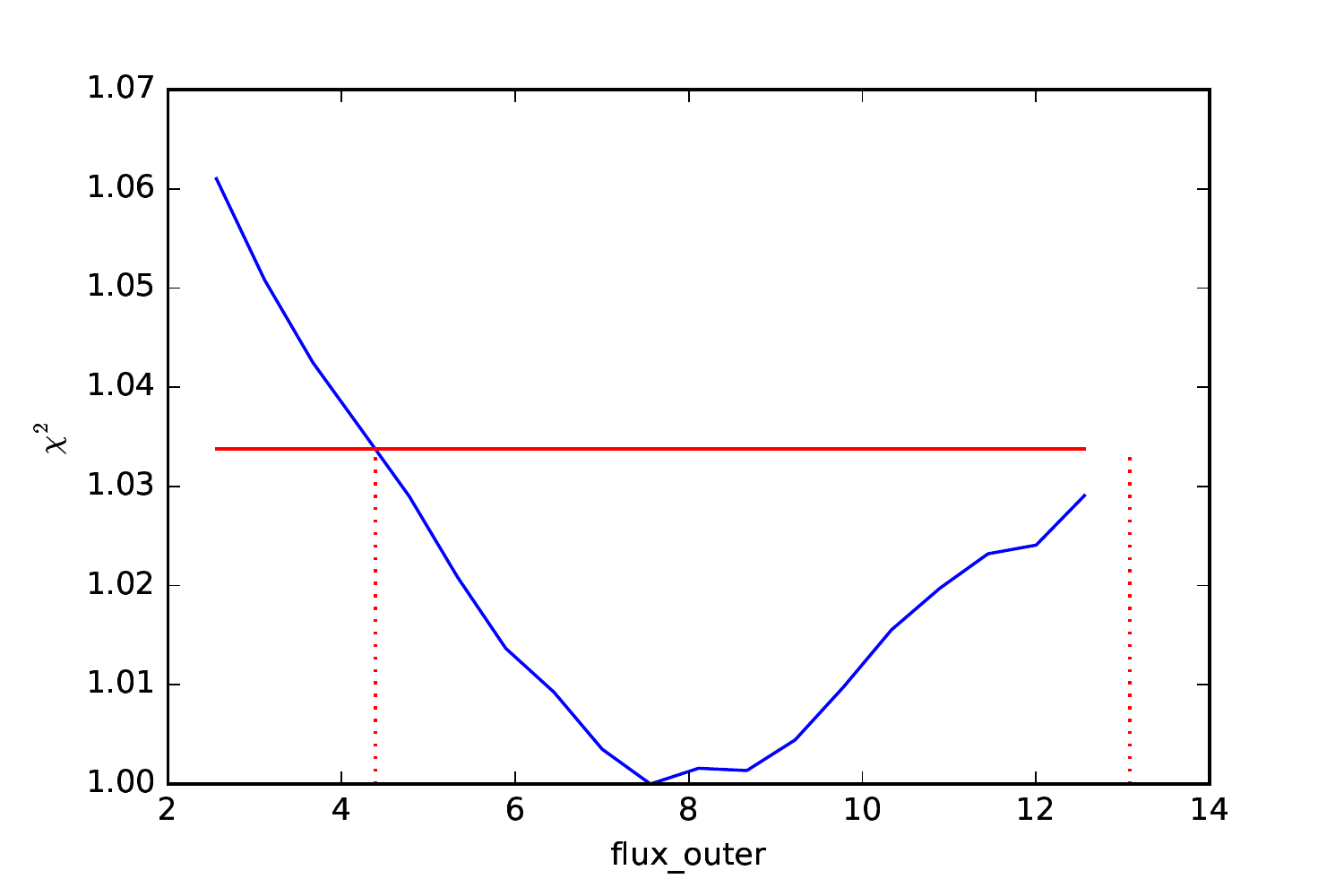} &  \includegraphics[width=4cm]{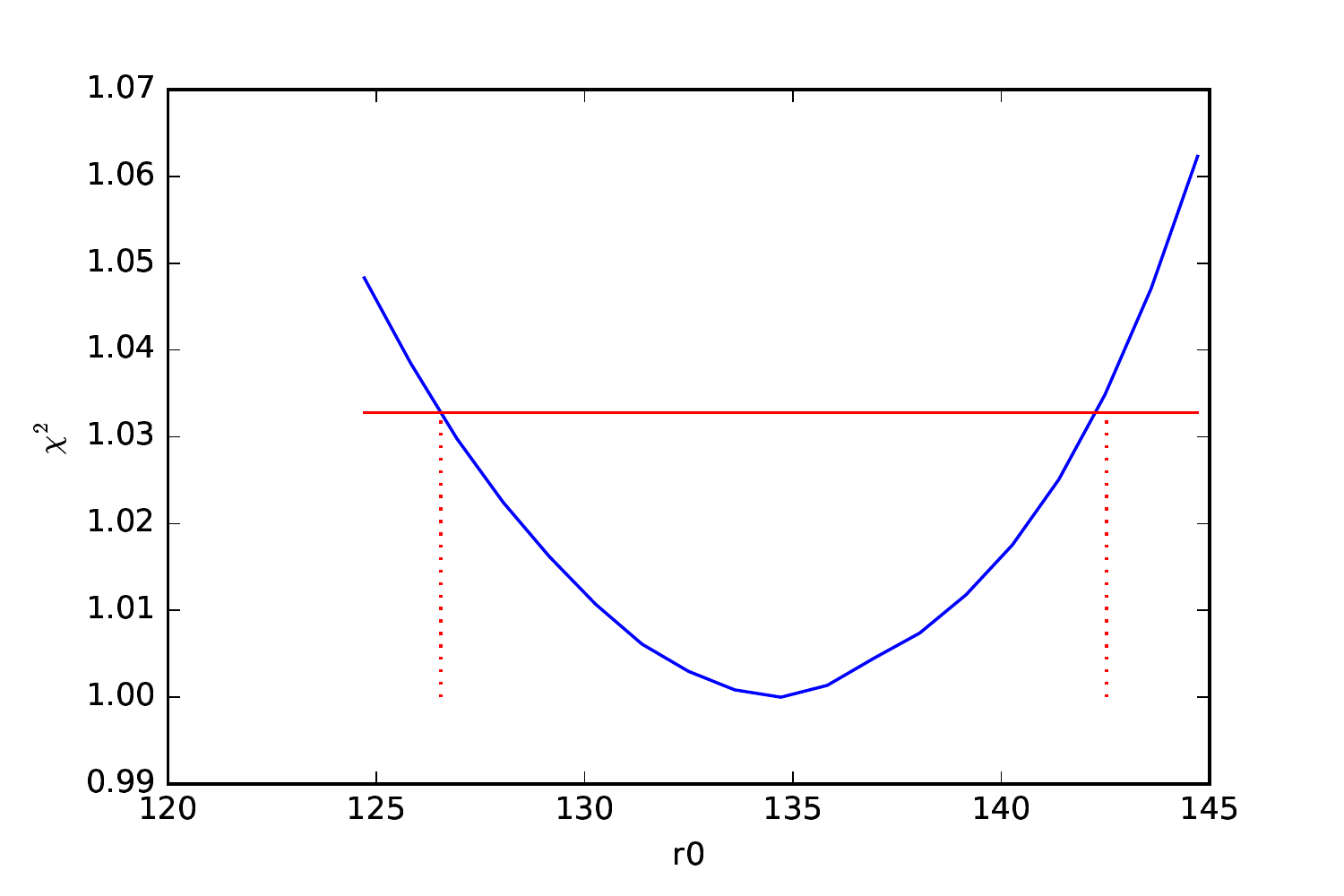} \\	
\end{tabular}
\caption{Minimization of the free parameters for the model of a single belt and a halo (left column) and of the model with two belts (right column).}
\label{fig:tworinghal}
\end{center}
\end{figure}

The Fig. \ref{fig:modsing} show the $\chi^{2}$ minimization for each of the five free parameters of the single ring model (\S~\ref{sec_inner_ring}). The Figure \ref{fig:tworinghal} show the  $\chi^{2}$ minimization for the free parameters of the models with two belts, or one belt and a halo (\S~\ref{sec_outer_ring}). The 1$\sigma$ $\chi^{2}$ levels used to estimate the error bars on each parameters are reported in red. 

\section{Detected point sources and limits}
\label{section:companions}
We detect 10 candidate companions (CC) at large separations in the IRDIS H2 and H3 images. Their astrometry and predicted mass -- assuming they are bound --  are reported in Tab. \ref{tab:CC}. All those point sources lie outside of the field-of-view of the IFS. None of them is aligned with the disk's position angle.  

\begin{figure}
\begin{center}
\includegraphics[width=8cm]{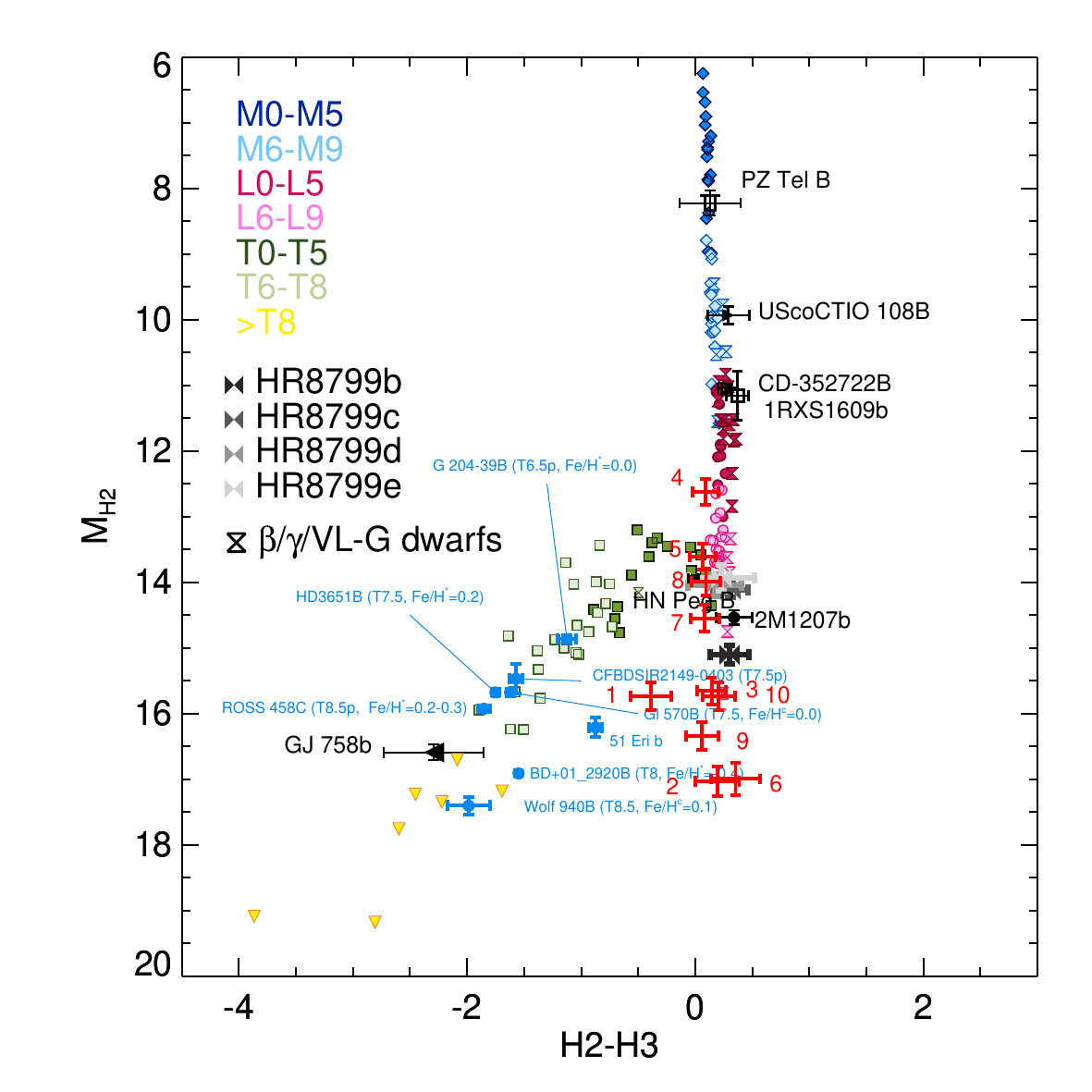}
\caption{Placement of the candidate companions from Tab. \ref{tab:CC} into color-magnitude diagrams.}
\label{fig:CMD}
\end{center}
\end{figure}

The placement of the candidates in color-magnitude diagrams can help to determine if they have both the luminosity and colors typical of cold substellar companions,  or if they are rather background objects. Such diagrams have been created for the IRDIS filters \citep[see a description in][]{2016arXiv160606654M}. We show the placement of the CCs of HIP~67497 in Figure \ref{fig:CMD}. The CCs \#2, 6, and 9 have the luminosities typical of T dwarfs but the colors of MK dwarfs, the most common contaminants. This indicates that these CCs are likely to be background objects. All the remaining  CCs, but the CC \#1 and 2, are retrieved  in the HST/STIS (optical) images of \cite{2016IAUS..314..175P} and are therefore  likely background stars.

\begin{figure}
\begin{center}
\begin{tabular}{c}
\includegraphics[width=8cm]{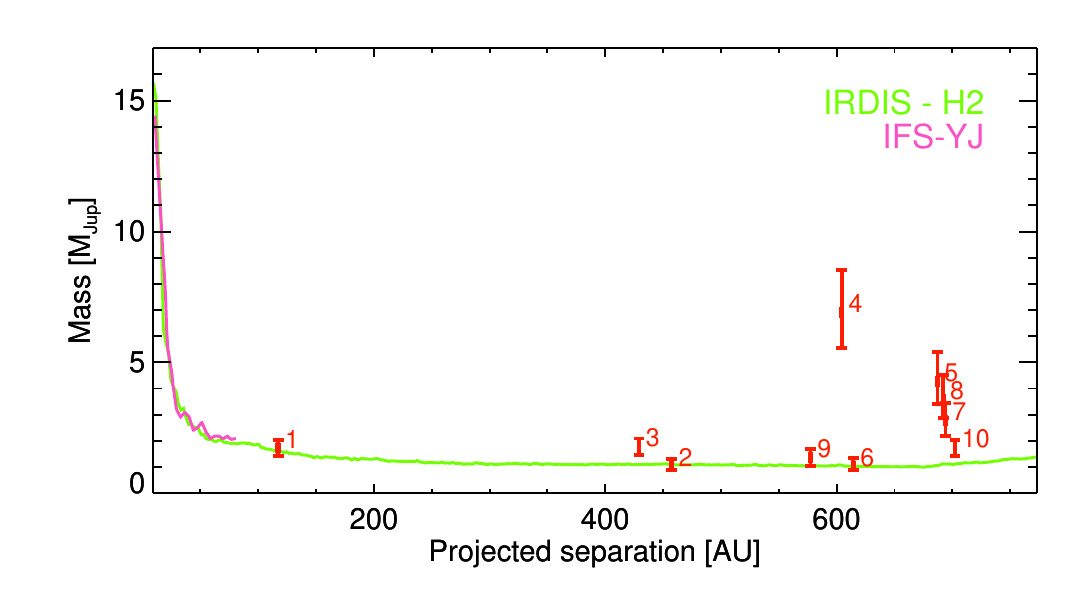} \\
\includegraphics[width=8cm]{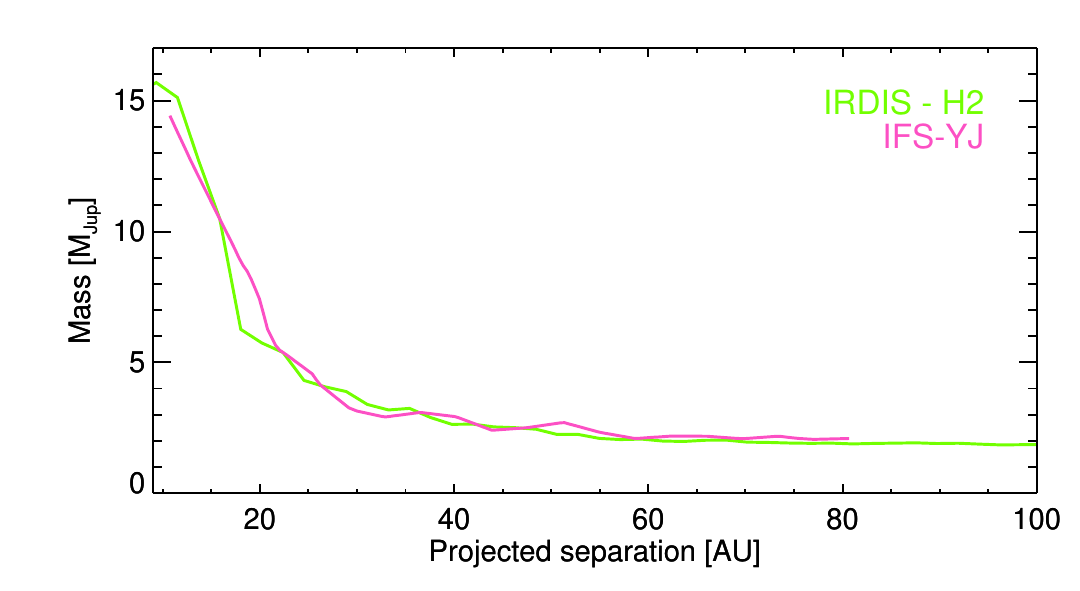} \\
\end{tabular}
\caption{Detection limits (5$\sigma$) for the IFS (spectral PCA) and IRDIS (TLOCI) converted to mass. The candidate companions are reported in red. Top: full separation range. Bottom: zoom at inner working angles.}
\label{fig:dl}
\end{center}
\end{figure}

 We estimated the detection limits  of the IFS  via the injection of fake companions with flat spectra into the datacubes and used the COND evolutionary tracks to convert the derived contrasts to masses \citep{2003A&A...402..701B}. The IRDIS detection limits were estimated from the TLOCI coefficients and the local level of the noise. We report those limits in Fig. \ref{fig:dl}. The data are sensitive to planetary-mass companions down to 15 au.

\label{AppB}
\begin{table}[t]
\caption{Photometry and astrometry of point sources. The predicted mass if bound to the star is reported into the last column.}
\label{tab:CC}
\begin{center}
\begin{tabular}{llllll}
\hline\hline
\#		&		Band			&	Contrast &	 PA & Sep	 &$\mathrm{M_{Pred}}$ \\
		&						&				 & ($^{\circ}$) & (mas)	&	($\mathrm{M_{Jup}}$) \\
\hline
1					&			H2		&	13.33$\pm$0.09	&	185.5$\pm$0.3 & 1092$\pm$6 & 2$\pm$1 \\	
					&			H3		&	13.71$\pm$0.13 & 185.3$\pm$0.4 & 1088$\pm$8 &  2$\pm$1 \\
2					&			H2		&	14.62$\pm$0.12	&	71.9$\pm$0.1	&	4258$\pm$7 & 1$\pm$1 \\
					&			H3		&	14.42$\pm$0.13 &	71.9$\pm$0.1  &	4264$\pm$7 &  1$\pm$1\\
3					&			H2		&	13.24$\pm$0.08 & 293.3$\pm$0.1	&	3996$\pm$3 &  2$\pm$1 \\
					&			H3		&	13.10$\pm$0.07 & 	293.3$\pm$0.1	&	3998$\pm$3 & 2$\pm$1\\
4					&			H2		&	10.21$\pm$0.07 &  100.4$\pm$0.1	&	5626$\pm$2 & 7$\pm$2\\
					&			H3		&	10.12$\pm$0.06 & 100.4$\pm$0.1	&	5628$\pm$2 & 7$\pm$2 \\					
5					&			H2		&	11.20$\pm$0.07	& 155.8$\pm$0.1	&	6398$\pm$2 & 4$\pm$1\\
					&			H3		&	11.13$\pm$0.07	&	155.8$\pm$0.1	&	6400$\pm$3 & 4$\pm$1\\	
6					&			H2		&	14.58$\pm$0.16	&	279.9$\pm$0.1	&	5721$\pm$6 & 1$\pm$1\\
					&			H3		&	14.12$\pm$0.12	&	280.0$\pm$0.1	&	5724$\pm$7 & 1$\pm$1\\
7					&			H2		&	12.13$\pm$0.07	&	285.6$\pm$0.1	&	6462$\pm$3 & 3$\pm$1\\
					&			H3		&	12.05$\pm$0.08	&	285.7$\pm$0.1	&	6458$\pm$3 & 3$\pm$1\\
8					&			H2		&	11.58$\pm$0.07	&	283.7$\pm$0.1	&	6444$\pm$3 & 4$\pm$1\\
					&			H3		&	11.48$\pm$0.07 & 	283.7$\pm$0.1	&  6439$\pm$3 & 4$\pm$1\\
9					&			H2		&	13.92$\pm$0.09	&	351.8$\pm$0.1	&	5376$\pm$4 & 2$\pm$1\\
					&			H3		&	13.87$\pm$0.09	&	351.8$\pm$0.1	&	5375$\pm$5 & 2$\pm$1\\	
10					&			H2		&  13.32$\pm$0.09 & 342.6$\pm$0.1	&	6539$\pm$5 & 2$\pm$1\\
					&			H3		&	13.12$\pm$0.08 & 342.6$\pm$0.1	&	6541$\pm$4 & 2$\pm$1\\																							
\hline
\end{tabular}
\end{center}
\end{table}

\section{Putative perturbers}
\label{AppC:perturbers}
We investigate the presence of one, two and three giant planets as responsible for the gap between the two possible belts resolved around HIP~67497 using numerical simulations based on published analytical formulaes. A detailed description will be made in Lazzoni et al. 2016 in prep.

If we make the assumption that one planet on a circular orbit only is responsible for the gap, we find that this planet should have a mass of 20.6 M$_{Jup}$ and semi-major axis a=90.5 au following  \cite{2015ApJ...799...41M}. For the case of an eccentric planet, we used the expressions for the half-width of the chaotic zone $\Delta a$ found by \cite{1980AJ.....85.1122W}  (for planet eccentricity $\leq$0.3) and \cite{2012MNRAS.419.3074M} (for eccentricity  $>$0.3)  as a basis: 

\begin{equation}
\label{eq:1}
\Delta a = 1.3 \times (M_{p}/M_{Star})^{2/7}\times a_{p},
\end{equation}

\begin{equation}
\label{eq:2}
\Delta a = 1.8 \times (M_{p}/M_{Star})^{1/5}\times e^{1/5} \times a_{p},
\end{equation}

with $M_{p}$ the planet mass,  $M_{star}$ the primary star mass, and $e$ the eccentricity of the disk. In our analysis, we consider the planet to arrive at periastron in the nearest point to the inner belt and at apoastron in the nearest point to the outer one. Then we substitute in the previous expressions for  $\Delta a$  the value of the semi-major axis with the positions of periastron and apoastron in turns. In that case, we predict that planets with low eccentricities ($e=0.13$) can have masses below our detection threshold.

We also investigated the case of 2 or 3 equal-mass planets on circular orbits. We considered planets as near as possible to each others \citep[maximum packing,][]{1993Icar..106..247G} in order to have yet a stable system but, at the same time, to obtain completely chaotic regions between the planets such that the disk material could not survive in it. For two planets on circular orbits, we followed the results of \cite{2014MNRAS.442.1110M} and found that 2.7 M$_{Jup}$ equal-mass perturbers  with semi-major axis  of 73  and 105 au could clear the gap. For the case of a three planet systems with circular orbits, the relations of  \cite{2014MNRAS.442.1110M}  give perturber masses of 0.14 M$_{Jup}$  and semi-major axis of 64.4, 87.5,  and 118.7 au. We considered a last configuration with two equal-mass planets on eccentric orbits, supposing again the maximum packing condition, and following the equations \ref{eq:1} and  \ref{eq:2}. We find that low masses (0.1-0.4 M$_{Jup}$) with quite small eccentricities are needed to dig the entire gap.

We recall that multiple  assumptions are made here (equal mass pertubers, maximum packing). Therefore, we  believe that a more in depth  dynamical analysis of this system is requiered to understand the morphology of the disk (N-body simulations; beyond the scope of this letter).


\end{appendix}

%
%

\end{document}